# Review of oxygen measurement and relevance to the mechanisms of FLASH radiotherapy

**Brian W. Pogue[1,2,3], David I. Hunter[1], Corbin Narita[1], William S. Thomas[3], Xu Cao[1], Harold M. Swartz[1,2]**

[1]Thayer School of Engineering at Dartmouth, Hanover, NH, USA 03755

[2]Giesel School of Medicine at Dartmouth, Hanover, NH, USA 03755

[3]Department of Medical Physics, University of Wisconsin School of Medicine and Public Health, Madison, WI 53705

## Abstract

FLASH radiotherapy (FLASH-RT) is the phenomenon of relative sparing of normal tissue when ultra-high dose rates (UHDR) are used compared with conventional dose rates (CDR) as clinically used. Despite extensive investigation, the underlying mechanisms remain unexplained. Among the proposed hypotheses, tissue oxygen has consistently been a central theme because oxygen is the most dominant factor known to modulate radiation-induced damage. The factors implicated in FLASH sparing include the baseline partial pressure of oxygen ($pO_2$), transient radiolytic oxygen consumption (ROC), and oxygen-dependent changes in the chemistry of reactive oxygen species (ROS) that vary with dose rate. This review synthesizes current evidence on *in vivo* oxygen measurement techniques, highlighting their capabilities and limitations in capturing the spatial and temporal heterogeneity of tissue oxygenation. Key experimental studies in skin are summarized and interpreted by modulating oxygen levels via changes in inspired oxygen gas and vascular clamping interventions, demonstrating that the FLASH effect occurs only at intermediate baseline $pO_2$ (normoxic or slightly hypoxic) values, but not at hypoxia or hyperoxia. Direct measurement of oxygen consumption during UHDR irradiation is possible, providing one of the first in situ measurements of radiation chemistry in patients. In parallel, recent advances in fast in vitro radiation chemistry assays indicate that UHDR irradiation alters radical yields, favoring increased production of solvated electrons and reduced hydroxyl radical–mediated damage. Taken together, the available data suggest that the FLASH sparing effect arises from an interplay among the delivered dose and dose rate, local oxygen availability, and radiation chemistry, with tissue-specific variation in scavenging, leading to altered biological responses across the CDR-to-UHDR shift. This more complex interpretation seems more likely than the simpler interpretation of broad-area radiolytic oxygen depletion alone. However, it must be acknowledged that we have partial data on all aspects of this hypothesis, and further improvements in oxygen sampling are very likely to help in understanding ROS and scavenging effects *in vivo*. Key challenges in quantifying oxygen dynamics *in vivo* are highlighted, and the conclusions are used to identify critical areas for future research to enable mechanistic understanding and clinical translation of FLASH-RT.

## 1.0 Introduction

The normal tissue sparing effect observed in FLASH radiotherapy (FLASH RT), delivered at ultra-high dose rates (UHDR), is intrinsically linked to oxygen-dependent processes at multiple hierarchical levels, from radiochemistry to tissue response [1–3]. Despite intensive investigation across numerous laboratories, the causal mechanisms responsible for FLASH-mediated normal tissue protection remain unresolved. Nevertheless, a growing body of experimental evidence indicates that UHDR irradiation alters reactive oxygen species (ROS) production, oxygen consumption, and oxygen-mediated toxicity compared with conventional dose-rate (CDR) irradiation [1–6]. The central role of oxygen in modulating radiation-induced cellular and tissue injury is well established, and its potential contribution to the FLASH effect—particularly through transient oxygen depletion and altered radical kinetics—has been the subject of recent focused reviews that synthesize original experimental findings [7]. In this context, accurate characterization of oxygen dynamics is critical for the mechanistic interpretation of FLASH-RT studies. This review, therefore, focuses on the core principles and practical challenges of *in vivo* oxygen measurement in the setting of UHDR irradiation, emphasizing how measurement technique, spatial and temporal resolution, and biological perturbation influence observed oxygen levels and their interpretation. We systematically examine key methodologies for quantifying tissue oxygenation, assess their respective strengths and limitations, and discuss how these factors constrain current conclusions about FLASH mechanisms. By linking oxygen measurement strategies to radiochemical and biological endpoints, this review aims to identify critical gaps and inform future experimental designs needed to more definitively elucidate the role of oxygen in both CDR and UHDR delivered radiotherapy research.

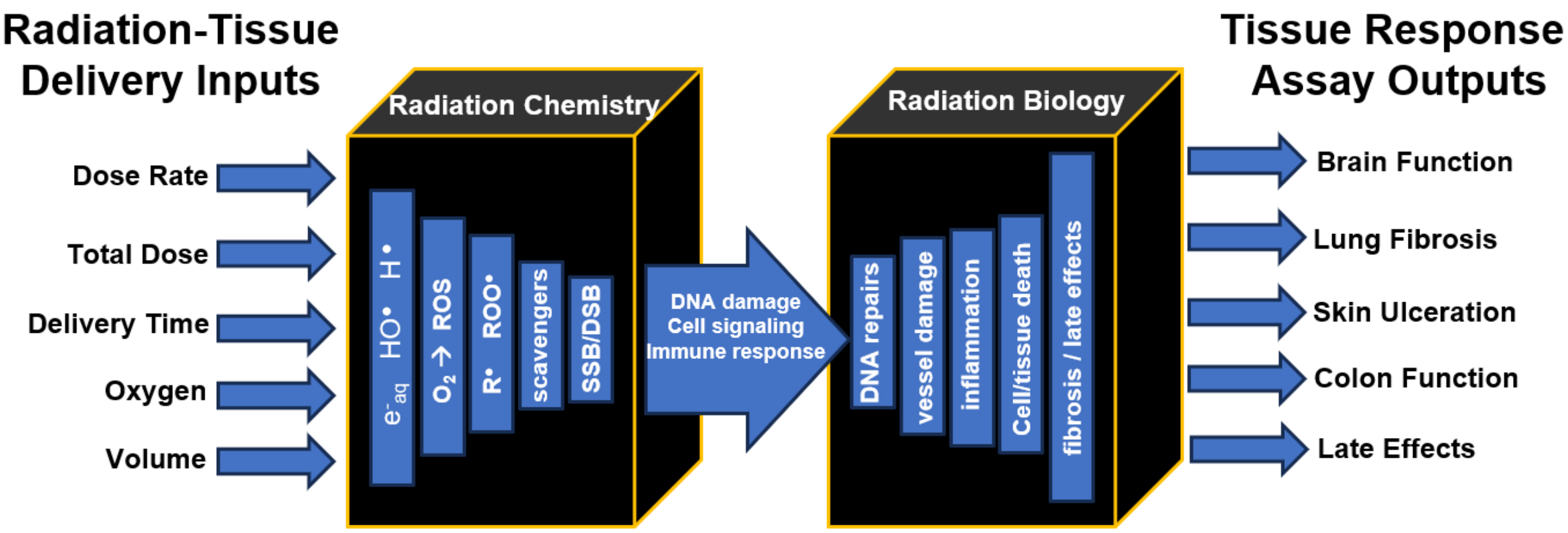

**Figure 1.** A conceptual framework for how input radiation delivery parameters (left arrows) influence the output of tissue organ response assays (right arrows) is illustrated. This framework is designed to consider the temporal cascade of radiation-chemistry factors that transition from hydrolysis-derived primary radicals through ROS formation and peroxyl radicals to radical scavenging, ultimately culminating in DNA damage as single- and double-strand breaks (SSBs & DSBs). These lead to radiation-induced biological responses, including DNA repair and endothelial cell damage that induces microvessel injury, all contributing to the inflammatory cascade. The complexity of cell infiltration and death that occur within days to weeks after FLASH appears to be altered as well, and, finally, late effects such as fibrosis and loss of tissue function constitute the catastrophic damage that ultimately limits the dose to normal tissues in all radiotherapy planning. how FLASH studies are carried out, with the major unknown black boxes illustrated by cascading stages of radiation chemistry and radiation biology.

Much of what is known in FLASH-RT research has been developed through systematic experiments in which one input delivery parameter is varied at a time, and the output response is measured in a biological function assay [8,9], as illustrated conceptually in **Figure 1**. There are two major black boxes involved in the process: radiation chemistry and radiation biology, either of which could be the key mechanism underlying the FLASH effect, or an interweaving mechanism could link them. After water radiolysis into primary radicals, the dominant radicals that are thought to influence DNA damage are hydroxyl radicals (HO•), aqueous or solvated electrons ($e^-_{aq}$), and atomic hydrogen (H•). Molecular oxygen ($O_2$) plays a central role in radiation chemistry, giving rise to a complex cascade of reactive oxygen species (ROS), including superoxide ($O_2$•-), hydrogen peroxide ($H_2O_2$), and singlet oxygen ($^1O_2$) [10]. Biochemical scavengers for hydroxyl radicals, such as glutathione (GSH), ascorbate, uric acid, and melatonin, can significantly reduce damage, and ROS can be suppressed by the activity of enzymes, including superoxide dismutase (SOD), catalase, glutathione peroxidase, and peroxiredoxins [11–14]. Studies of oxygen-mediated damage to nucleic acids are well known from direct *in vitro* work demonstrating the effects of single-strand breaks (SSBs) and double-strand breaks (DSBs) in DNA [15–20]. In vivo, numerous correlational studies have reported similar effects. The link to biological damage is classically through DNA damage. Still, it can also involve cellular proteins and changes in signaling pathways, as well as systemic factors such as immune modulation or response. The biological cascade is inherently complex and can be linked to oxygen at multiple levels, with DNA repair being a primary area of study [19]. But the damage from radiotherapy at high dose levels is known to induce microvascular damage that can be both acute and chronic. There is documentation of inflammatory reactions to large doses, which can induce fibrosis, tissue damage, and cell death [15][21–23]. Ultimately, many of the microscopic events are not easily assayed, so much of the FLASH data has been presented as macroscopic tissue responses, shown at right in Figure 1, including various assays of brain function, lung fibrosis, skin ulceration, colon crypt morphology, and late effects on tissue function. Peering inside these black boxes can be done through strategic, hypothesis-driven experiments and computer simulations.

Molecular oxygen has a profound and fundamental effect on radiotherapy, yet paradoxically, tissue oxygen is largely ignored in clinical practice due to the challenges of comprehensive, accurate, and actionable measurement. This is in part due to the lack of clear in vivo data, compared with clearer in vitro data showing a strong oxygen dependence of radiosensitivity. A core concept in radiobiology, derived from cell studies, is the oxygen enhancement ratio (OER), which consistently shows that oxygen levels significantly influence radiation-induced cell death. Irradiating cells under either anoxia ($pO2 \simeq 0$ mmHg) or normoxia ($pO2 \simeq 30$-60 mmHg) can change the probability of cell death by a factor of 2.7 [117]. Hypoxia in tumors is correlated with radiation resistance, and numerous studies have shown that clamping off blood flow in tumors induces significantly less radiation damage at CDR [24]. While many correlated biological factors are induced in anoxic or hypoxic tissue, the absence of oxygen is known to reduce oxidative DNA damage, thereby leading to cellular death [25–27]. The hypothesized mechanism is that oxygen fixes DNA damage by binding to DNA radicals, thereby converting them into less fixable, longer-lived peroxyl radicals [27–29]. The lack of oxygen allows DNA radicals to be repaired more readily by thiols such as GSH; therefore, the overall observed effect of radiation damage may be less when measured by cellular survival.

It is important to recognize that oxygen is the largest external factor affecting radiotherapy-induced tissue damage, as it modulates toxicity more than twice the magnitude of the FLASH effect, as measured by a dose-modifying factor (DMF). This concept is illustrated in Figure 2a. In data from Hansen et al. (2026) [30], the skin-damage morbidity in mice shifts from an LD50 of 31 Gy with UHDR to 45 Gy and to 64 Gy in the absence of oxygen. However, the extent of the FLASH sparing effect can be modulated by oxygen availability, as observed in several studies [1–4,31], as illustrated by the data in Figure 2b, adapted from Hunter et al. (2026) [116]. This paradigm of oxygen modulating the effect of UHDR relative to CDR will be examined in detail, with attention to experimental design and measurement fidelity.

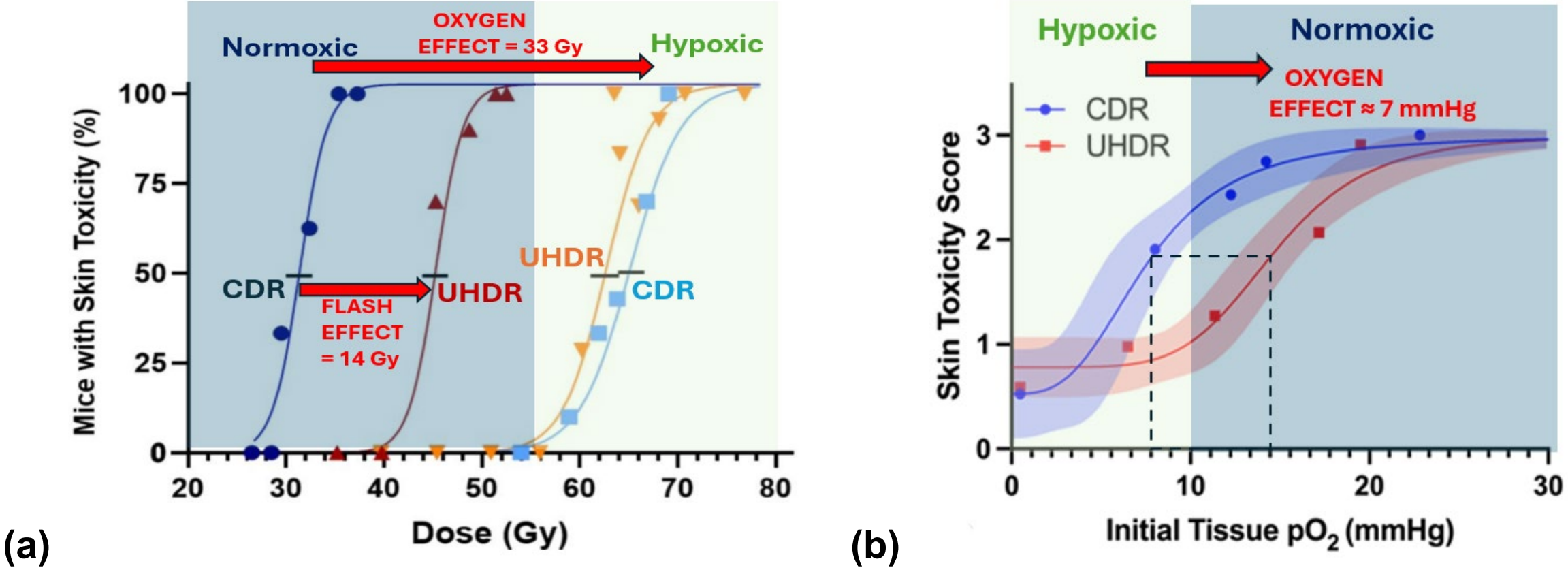

**Figure 2**. The presence of oxygen is known to affect radiotherapy-induced damage profoundly. An illustration of the extremity of this is shown in (a) above, adapted from [2,30], illustrating that the increased damage resistance from normoxia to complete hypoxia is 33 Gy for skin damage. In comparison, the increased damage resistance from conventional dose rate (CDR) to ultra-high dose rate (UHDR) is 14 Gy. In comparison, when tissue pO2 is varied, shown in (b), adapted from [116], there is an apparent shift of sensitivity to higher levels, indicating that lower pO2 is less sensitive to the radiation at UHDR.

To understand the role of oxygen in FLASH, it is important to start with a careful examination of how oxygen is measured, as this is part of why oxygen is largely ignored in clinical radiotherapy [32–35]. The measurement approach will yield different oxygen values depending on the sampling site and measurement method. Most attempts to measure tissue oxygen tend to yield highly volume-averaged values of oxygen present throughout a region [36–38]. These complicating factors are part of the logistical reasons why oxygen is often not measured or considered *in vivo*. Additionally, there are temporal changes in oxygen that are both physiological and radiobiological [39,40], and these have become an interesting and important part of understanding how dose rate affects radiation chemistry. Measurement and hypothesis testing based on transient oxygen changes are very useful tools for parsing out the underlying mechanism of oxygen in FLASH. This review details the measurements and hypotheses tested, and attempts to summarize key findings and possible future directions.

## 2.0 *In vivo* Oxygen Measurement Systems

### *2.1 Blood Oxygen Saturation ($SO_2$) Versus Tissue Partial Pressure of Oxygen ($pO_2$)*

Complete information on *in vivo* oxygenation is nearly impossible to quantify adequately for many reasons. It is heterogeneous at the microscopic level; it can vary rapidly within seconds, and its resupply from the blood depends on saturation and blood flow [36]. Because of these issues and limitations in measurement techniques, the tools used to characterize tissue oxygen levels routinely do not fully represent the true value. There are various methods to quantify the partial pressure of oxygen (pO2) *in vivo*, ranging from invasive electrodes and fiber-optic probes to molecular reporters and surface electrodes (see **Figure 3a**). Measurements of blood oxygen saturation (SpO2) are used much more widely due to their relative ease of acquisition. Still, the value of $SO_2$ is nonlinearly related to tissue pO2, via the hemoglobin binding rate and many other biochemical factors, such as pH and $CO_2$. As a result, SO2 is not uniquely linked to the available oxygen at the tissue level, particularly for biological and radiobiological purposes. Thus, in radiation biology, the most useful information requires direct measurement of pO2 to assess its role in ROS generation quantitatively. In **Figure 3b**, the relative rankings of the major $pO_2$ measurement systems are arranged on a radar plot, roughly assessing their capabilities on 5 major axes that assess their (i) signal-to-noise ratio (SNR), (ii) acquisition time, (iii) smallness of volume of tissue sampled, (iv) *in vivo* location from large to intracellular, and (v) their level of non-invasiveness. These are roughly ranked, with larger plot areas indicating higher potential value for the system.

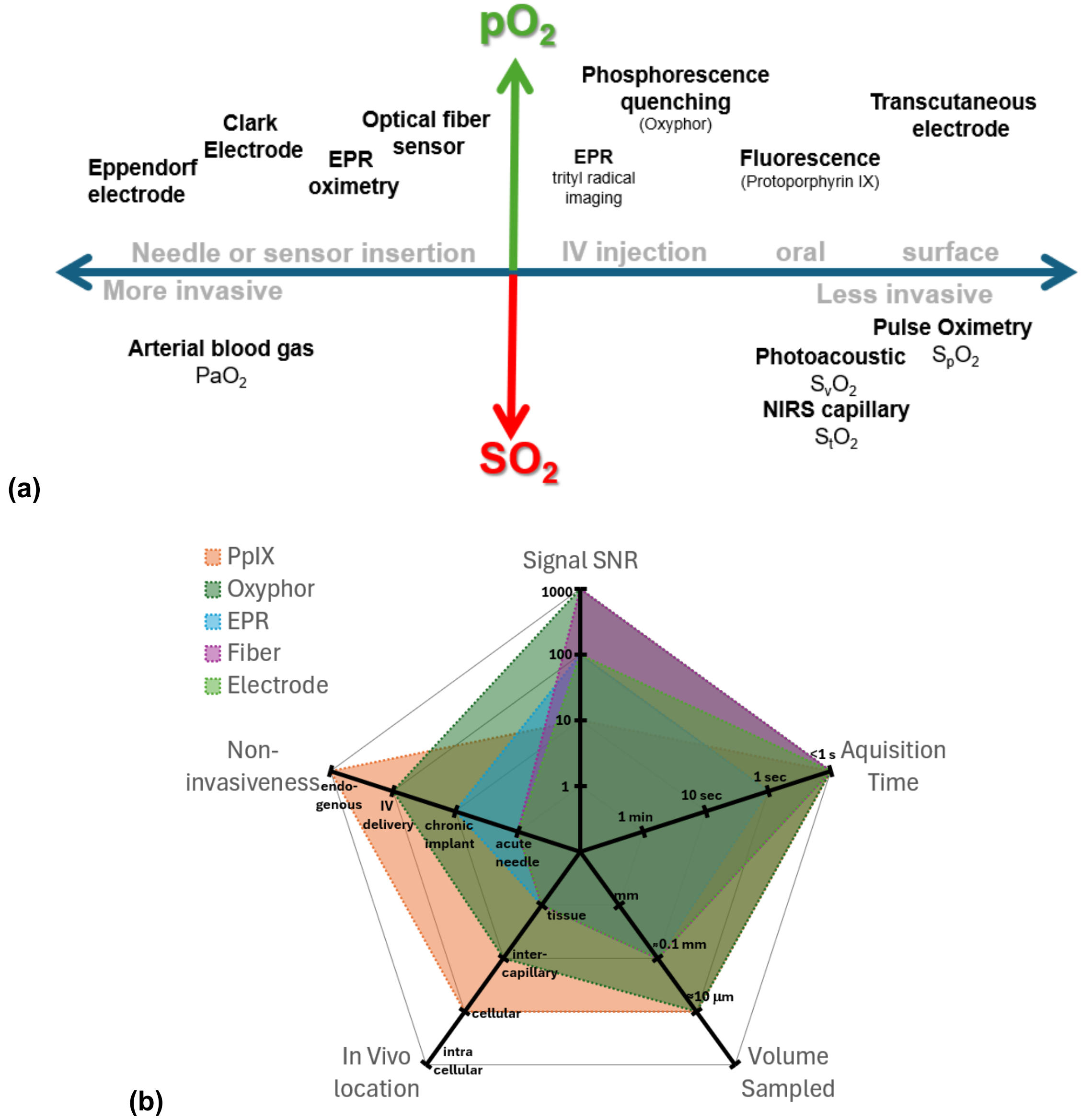

**Figure 3.** Illustration (a) of the oxygen partial pressure ($pO_2$) and blood oxygen saturation ($SO_2$) measurement techniques possible, arranged on a level of invasiveness; (b) of the 5 primary *in vivo* oxygen $pO_2$ measurement devices and their benefits in terms of relative (i) SNR, (ii) lower acquisition time, (iii) smallness of volume of tissue sampled, (iv) *in vivo* location in terms of intracellular, extracellular or large volume, and (v) level of non-invasiveness. Molecular reporters are the least invasive and most reliable for radiation-chemistry measurements and fast reporting.

Beyond measuring bulk $pO_2$ values, the tissue compartment (i.e., vessels, interstitial fluid, cells) in which it is measured can significantly affect the interpretation of the data, because $pO_2$ varies from high values (near 80-100 mmHg) in arterial blood to 30-40 mmHg in venous blood. In contrast to $pO_2$, $SO_2$ is often used as a surrogate for oxygenation, even though it only indirectly reflects tissue oxygenation. $SO_2$ is near 100% in arteries, drops to lower levels in capillaries, the primary site of outward oxygen diffusion, and remains nearly constant in veins. Normal humans would have an arterial oxygen saturation ($SaO_2$) of 90-100% [41], while tissue capillary $StO_2$ can be as low as ≈ 60% [42]. The non-linear binding affinity of oxygen to hemoglobin is optimal for this delivery process. Still, because of the spatial distribution impacted by binding-convection-diffusion, the relationship between $pO_2$ and $SO_2$ is never linear. Because most $pO_2$ measurements are macroscopic, the reported value is highly averaged over the volume near the reporter, mixing arteriole, capillary, venule, and tissue values while being biased toward the location of the sensor probe [43].

### *2.2 Probes of $pO_2$ — Invasive Versus Minimally Invasive Versus Molecular Reporters*

Most inserted probes, such as electrodes, optical fibers, or solid paramagnetic probes, are large (on the order of millimeters) compared with intercapillary distances (100-300 microns). So the insertion of these by the needle can cause tissue trauma and alter interstitial pressure, ultimately affecting oxygenation [33,44]. These can be stably inserted when carefully placed for electrodes or fibers, and can be minimally invasive when designed with a small diameter (100s of microns) [45]. However, there are two major issues with these approaches. First, an inflammatory reaction to the insertion/object can be unpredictable, and secondly, the measurement time is typically several seconds to minutes to obtain a stable reading. The most complete and thereby reliable data on human tissue $pO_2$ were obtained using electrodes inserted through tissue over several centimeters [34,35,38,39,46], in a series of tumor studies conducted in the 1990s with the Eppendorf electrode system [44,47]. The process of moving the needle and measuring tissue volumes yielded a histogram of values quantifying $pO_2$ heterogeneity within the tissue. This system is no longer sold or available for human use, but individual Clarke-type electrodes are used in preclinical work for fixed-location scientific measurement [48]. It is important to appreciate that because oxygen diffusion and capillary blood flow are affected by nearly all pressures in tissue, the actual insertion and any potential inflammation around it, can significantly change the very pO2 that is attempted to be measured, and so insertion probes can routinely report lower or higher pO2 based upon pressure or inflammation. Smaller probes typically have faster response times and can be less perturbative of the oxygen field in tissue, though they also suffer from higher noise levels [49]. Finally, the temporal response of these probes and their resistance to high-radiation environments are major issues that can limit their utility for fast or in-field measurements in the radiation beam. Thicker needles have a slow response time of seconds, whereas smaller needles can have a sub-second response time, albeit with higher electrical noise and greater biological variability. Similarly, optical fiber probes or electron paramagnetic resonance (EPR) probes may have response times of several seconds to minutes [36,50] to equilibrate with the surrounding environment, so measurements of fast pO2 transients may be limited by these systems' temporal response times.

In contrast, molecular reporters can diffuse from capillaries into tissue, providing a distributed measurement of tissue pO2 [51,52]. Their chemical design dominates their partitioning, pharmacokinetics, and microscopic localization, which in turn determine where $pO_2$ is actually measured within the tissue. The most effective in vivo molecular reporters use oxygen-quenching luminescence, such that their emission-lifetime measurements provide an intensity-independent assay of local $pO_2$. Alternatively, EPR has been used extensively with molecular probes, in which oxygen-induced collision-induced linewidth broadening can be used to quantify $pO_2$ [53]. The two most widely used luminescent optical sensors are based on metalloporphyrins [51,52,54,55] or ruthenium complexes [56–58], each with multiple formulations to protect the chromophore and to modulate biocompatibility and localization. A more recent discovery has been protoporphyrin IX delayed fluorescence, a quenching-based reporter of tissue oxygen [59–61]. Each has its own strengths and weaknesses, depending on localization, pharmacokinetics, signal strength, lifetime, photostability, and biocompatibility.

### *2.3 What pO2 Measurements are Needed to Follow Acute Changes in FLASH-RT?*

For a FLASH radiation therapy mechanism study, molecular tissue oxygen reporters have a significant advantage in being distributed throughout the tissue volume and in providing fast, high-SNR measurements unaffected by the radiation. Oxyphor molecules, in particular, are a class of biocompatible probes composed of metalloporphyrins within a pegylated dendrimer shell [62–64], that can be administered many ways *in vivo* (intravenous, interstitial, oral) and provide an extracellular measurement of oxygen that is throughout the volume being sampled at kHz rates by a phosphorescence lifetime fiber probe, and is undisturbed by the radiation pulses when triggered appropriately. Alternatively, protoporphyrin IX is also measurable in a similar manner [59,65–67], and intracellular oxygen levels can be measured, although its calibration linearity and photostability are less robust than those of Oxyphors. Early work in UHDR irradiation studies has shown their value both in solution [68] and *in vivo* [69–71].

| **Key Points** | |
|---|---|
| **1** | Tissue oxygen $pO_2$ can be quantified by direct measurement, but has high spatial and temporal heterogeneity. All values have their bias, regional reporting or spatial averaging, and there is still potential for incomplete sampling. |
| **2** | Blood $SO_2$ parameters are more readily measured but do not have a predictable relationship to relevant tissue oxygenation. |
| **3** | Steady-state tissue pO2 can be measured in several ways (electrode, phosphorescence, fluorescence, fiberoptic, EPR). Repeatability varies with the level of invasiveness allowed to avoid disturbing the values. |
| **4** | Molecular reporters (phosphorescence and fluorescence) are ideal for many reasons — localization, non-perturbing, millisecond response time, lifetime-based, stability, fast transient responses before/after the UHDR delivery ($\Delta pO_2$) |
| **5** | Longer-term, repeatable, or chronically implantable measurements of pO2 may be useful for understanding chronic pathophysiological differences due to fractionation or microvascular changes. |

| **Open Questions and Future Directions** | |
|---|---|
| **1** | Measurement of site-specific oxygen at high temporal and spatial resolution will always remain a challenge, especially at sites distant from vasculature and intracellular compartments. |
| **2** | Microscopy imaging of oxygen and ROS in a radiation beam could help resolve this, although comparing oxygen and responses *in vitro* and *in vivo* is challenging. |
| **3** | Oxygen data should ideally be combined with downstream ROS measurements to determine the cascade of events and the interactive relationships with microenvironmental factors such as ROS scavengers, proteins, lipids, and nucleic acids. |

| 4 | Combining oxygen and ROS measurement with oxygen and ROS modeling may help bridge the gap in what cannot be measured *in vivo*. |
|---|---|

## 3.0 FLASH-RT Oxygen Measurement

### *3.1 In vivo Oxygen Measurement*

The high variability of tissue oxygen limits what can be determined experimentally. Baseline transient levels can vary within seconds, and dynamic homeostatic variations among animals and organs are regulated by factors such as hormones, baseline activity levels, and age. This baseline $pO_2$ is then further altered by temporal experimental effects of the animal preparation, such as anesthesia type, dosage or concentration, duration, and inspired ambient oxygen level, as well as conditioning factors such as body temperature and humidity. These factors commonly alter blood flow to each organ system, resulting in oxygen levels ranging from 0 to $pO_2 \approx$ 60-90 mmHg. Additionally, animals within the same cohort can exhibit baseline oxygen variations of up to ±50% [1]. Thus, the interpretation of oxygen in FLASH studies needs to take into account the biological variability and the measurement speed and accuracy of each measurement system.

Observed measurements of oxygen and how it affects the FLASH response can be separated roughly into categories, such as: (i) homeostatic levels from choice of species, sex, and organ systems, (ii) experimentally induced effects of steady state $pO_2$ value, and (iii) transient effects/observations of radiolytic oxygen consumption. Useful baseline oxygen measurements can be obtained with several slower measurement systems, although care must be taken to understand the absolute accuracy of each measurement. Commonly, that baseline pO2 is not measured in most homeostatic biological studies, with the assumption that all animals in a cohort have similar oxygen and oxygen effects; however, this assumption will be interpreted later and noted how inaccurate it can be.

In FLASH studies, in vivo oxygen measurements have been performed primarily in the skin due to its ease of access and stability during irradiation without surgical intervention. Tumors have also been measured, although care must be taken in interpreting the data, as measurements through the skin can potentially alter tumor measurements. The following table summarizes some of the most recent studies that included in vivo oxygen measurements in FLASH studies across different tissues, measurement techniques, and radiation modalities.

**Table 1:** Summary of *in vivo* oxygen experiments during FLASH radiotherapy.

| Technique | Species | Location | Modality | Conclusions | Ref |
|---|---|---|---|---|---|
| Mouse pulse oximeter | Rats | Blood | p | 70% $FiO_2$ causes deleterious effects in CDR and UHDR and hinders tumoral immune cell infiltration in CPT | [3] |
| EPR imaging, phosphorescence | Mice | Liver, Bowel | e | Toxicity proportional to Tissue $pO_2$. Females are more sensitive than males, potentially due to higher baseline $pO_2$. | [31] |
| Oxyphor | Mice | Normal tissue (flank) | e | ROC $g_{o2}$ *in vivo* is dependent upon baseline $pO_2$ | [72] |

| Oxyphor | Mice | Normal tissue (leg) | e | 100% $FiO_2$ negates the Flash effect. Female mice show increased radiosensitivity compared to males with UHDR but not CDR. | [4] |
|---|---|---|---|---|---|
| Oxyphor | Mice | Tumor, Leg flank | e | ROC is modest under UHDR, with normal tissue showing a greater decrease in oxygen than tumor tissue. | [73] |
| Oxyphor | Mice | Tumor, Leg | p | UHDR ROC is insufficient to cause tissue sparing. ROC correlated to baseline $pO_2$ | [74] |
| Oxyphor | Mice | Normal leg | p | ROC is insufficient for hypoxia, and $g_{o2}$ *in vivo* is dependent upon baseline $pO_2$ | [75] |
| Oxyphor | Mice | Tumor, Brain, skin, muscle | e | *In vivo* $g_{o2}$ is independent of tissue type | [76] |
| Oxyphor | Mice | Normal Tissue (leg) | e | FLASH effect is preserved if the total dose is delivered in <15 seconds. | [5] |
| Oxyphor, PpIX | Mice | Normal tissue (intracellular and extracellular) | e | Intracellular = extracellular ROC, and proportional to baseline | [64] |

### *3.2 Measurements and Interpretations of the Transient Oxygen Drop during FLASH-RT.*

The rapid decrease in tissue oxygen observed after UHDR irradiation is attributed to radiolytic consumption of oxygen via reactions that form long-lived intermediate radicals or covalent oxygen binding. Radiolytic oxygen consumption (ROC) or depletion is the absolute reduction in tissue pO2 during irradiation. ROC is thought to result from interactions with primary radicals generated by water radiolysis, as well as from interactions between reactive intermediates of ionized proteins and oxygen, forming peroxyl radicals, as illustrated in **Figure 4a.** Since ROC readily increases with the addition of protein to an irradiated solution [77], this suggests that the presence of oxygen enhances protein damage. The observed oxygen consumption depends on several factors, including baseline pO2, the total dose delivered, local microenvironment, and dose rate.

UHDR irradiation is typically delivered in microsecond pulses with repetition frequencies ranging from tens to hundreds of hertz, resulting in a total irradiation time on the order of milliseconds. However, the terminology used to describe dose rate often does not fully indicate whether the key parameter is the single-pulse dose rate or the dose rate averaged over a bunch of pulses. Given the assumption that the bunch of pulses dominates the average dose rate, the associated ROC occurs on an extremely rapid timescale and is effectively instantaneous from a biological perspective. As a result, resolving pulse-by-pulse oxygen transients remains challenging for Oxyphor phosphorescence probes, which generally require millisecond-scale acquisition times even in fast measurement configurations. Nevertheless, direct measurement of tissue oxygenation before and immediately after UHDR irradiation remains feasible. It provides a reliable quantification of ROC via the change in $pO_2$ ($\Delta pO_2$), defined as the difference between the initial $pO_2$ and the value measured immediately after irradiation.

Accurate assessment of $\Delta pO_2$ requires millisecond temporal resolution, as oxygen diffusion and vascular resupply can restore baseline levels within seconds following transient depletion. As shown in **Figure 4b,** UHDR irradiations typically have short delivery times ($\Delta t$ < 100 ms), and the oxygen diffusion timescale from capillaries out to the full tissue volume *in vivo* is about 5-10 seconds [78]. As a result, the abrupt oxygen consumption manifests as a rapid drop in $pO_2$ that occurs much faster than capillary resupply can restore, allowing detection as a net oxygen loss. Direct in vivo measurement of oxygen, however, can quantify $\Delta pO2$, the difference between the initial and final pO2 immediately after irradiation. If the measurement and total irradiation times are significantly faster than the diffusive oxygen resupply from the perfused capillaries, then $\Delta pO_2$ is a direct quantitative measure of ROC. However, using CDR that requires 10-30-second irradiations means that $\Delta pO2$ is dominated by biological oxygen resupply, and therefore it is not feasible to measure consumption *in vivo*. This can be seen in Figure 4C, where the quantification of $\Delta pO2$ is adversely affected by a non-instantaneous drop during irradiation, as the dose rate is not sufficiently high to cause a rapid change.

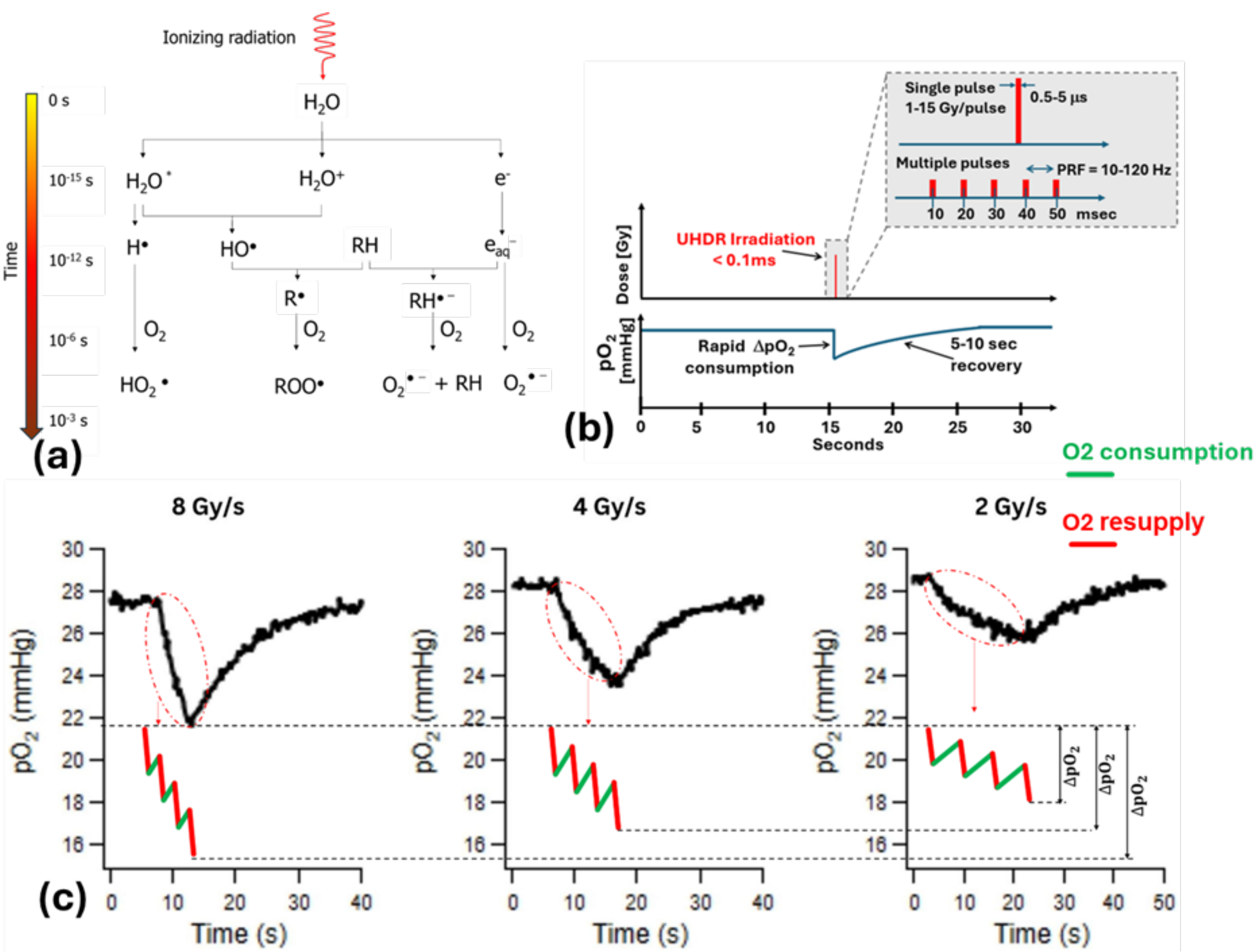

**Figure 4. (a)** The loss of oxygen comes from several locations in the complex ROS cascade, illustrated in a simplified diagram. (b) Illustration of the acute drop in oxygen, described as $\Delta pO_2$, with UHDR electron irradiation followed by oxygen recovery from diffusion in tissue. (c) *In vivo* measurements of $pO_2$ with Oxyphor, taken over time [76], illustrate the impact of dose rates on fast oxygen measurements: lower dose rates can yield a $\Delta pO2$ that is convolved with rediffusion, making quantification of the total $\Delta pO2$ unreliable or not representative of the total oxygen consumed. Color-line visuals depict times of radiolytic oxygen consumption (red), oxygen resupply (blue), and the final apparent $\Delta pO_2$. At higher dose rates, the measured $\Delta pO2$ is a direct measure of consumption because the resupply is less affected.

A brief note on the interpretation of dose rate is warranted here, as alluded to above. In electron FLASH studies, the mean dose rate is often reported relative to ROC; however, it is strongly influenced by the pulse repetition frequency (typically on the millisecond timescale). On the other hand, radiation chemistry is largely completed within nanoseconds. Given the pulsed nature of eFLASH irradiation, this temporal mismatch implies that the mean dose rate is not an adequate metric to interpret ROC results. Systematic reporting of all key dose-rate factors has been repeatedly recommended in several consensus reports and is necessary for interpreting data on the mechanism underlying FLASH. It is also very likely that the average dose rate over a bunch of pulses may be less predictive than the total irradiation time or the instantaneous dose rate per pulse. These issues need to be evaluated in further study.

### *3.3 The G-Value of Oxygen Consumption Measured In vitro and In vivo*

The first in vivo publications on oxygen consumption during UHDR irradiation appeared in 2021 [73,79], and only a few studies have built on this since. When the measured $\Delta pO_2$ reflects an absolute loss of oxygen, it can be normalized by the dose delivered, D, to estimate the radiochemical gO2 (i.e., g-value = $\Delta pO_2/D$). This is often reported in chemistry as the yield of molecules per 100 eV of irradiation, but in the medical physics literature as mmHg/Gy or μmol/Gy. Cao et al. [73] reported that UHDR irradiation consumed less oxygen than the equivalent dose of conventional radiation *in vitro* (g-value = 0.155 mmHg/Gy from UHDR and 0.19 mmHg/Gy from CDR, a nearly 20% difference). Other in-vitro studies report similar ranges, as listed in Table 2. Importantly, changes in g-value are not binary but rather monotonic with dose rate, and a more complete experimental interpretation of g-value in solution at varying dose rates has shown an inverse relationship with increasing dose rate [77,79].

**Table 2.** Tabulation of g-values for radiolytic oxygen consumption rates per unit Gray, for different solution types and radiation beam types at conventional dose rate (CDR) and ultra-high dose rate (UHDR).

| Study | Solution type | Modality | CDR (mmHg/Gy) | UHDR (mmHg/Gy) |
|---|---|---|---|---|
| Karle et al. | *In vitro* (BSA) | e | 0.35 | 0.32 |
| | | p | 0.29 | 0.28 |
| | | He | 0.25 | 0.23 |
| | | C | 0.19 | 0.18 |
| | | O | 0.18 | 0.16 |
| Sunnerberg et al. | *In vitro* (BSA) | e | 0.70 | 0.48 |
| Karle et al. | *In vitro* (BSA) | O | 0.18 | 0.16 |
| Thomas et al. | *In vitro* (BSA) | e | 0.56 | 0.32 |
| | | p | 0.45 | 0.35 |
| Cao et al. | *In vitro* (BSA) | e | 0.21 | 0.17 |
| El Khatib et al. | *In vitro* (BSA) | p | 0.55 | 0.49 |
| El Khatib et al. | Intracellular (HEK293T) | p | 0.31 | 0.32 |
| | CELL buffer | p | 0.48 | 0.39 |
| Van Slyke et al. | CELL buffer | p | 0.55 | 0.37 |
| Koch et al. | Phosphate | p | 0.25 | 0.29 |
| | H2SO4 | | 0.41 | 0.35 |
| | CELL/HEPES | | 0.85 | 0.40 |

* Maximal $O_2$ consumption i.e. > 30 mmHg plateau region

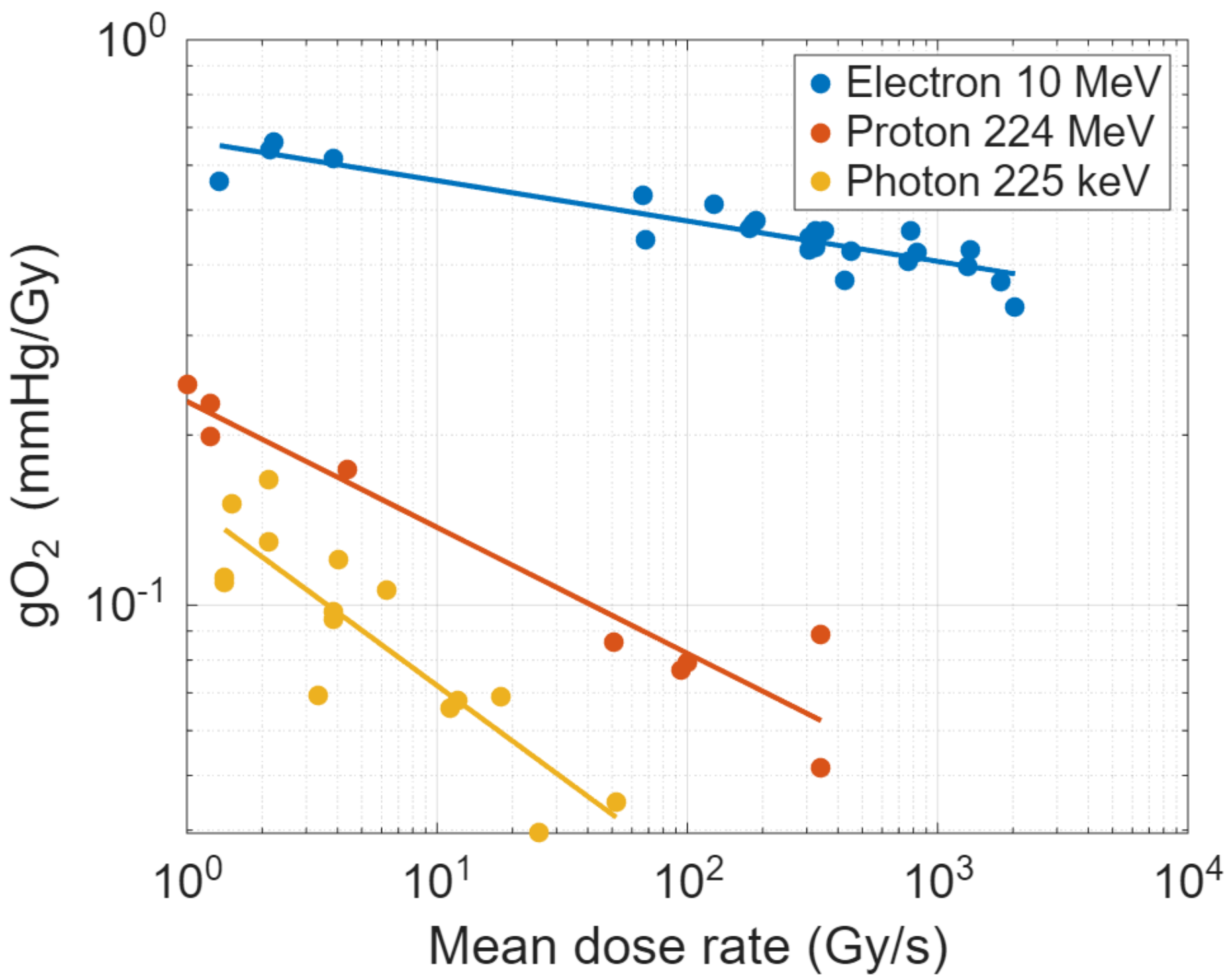

**Figure 5.** Data illustrating the oxygen consumption rate has an inverse relationship to incident dose rate, plotting oxygen consumption per unit dose (gO2 values) in water for 225 kV photon irradiation [79] and 224 MeV proton irradiation [79], and in 4% albumin aqueous solution for 10 MeV electron irradiation [77].

In an exhaustive *in vitro* study of radiolytic oxygen consumption variation with chemical environment, Koch et al [80] showed that intracellular reducing agents (i.e., NADH, glutathione, cysteine) can dramatically increase g-value up to 10x baseline values, while others (i.e., ascorbate and uric acid) decrease g-value by a factor of 2-3x and impose strong oxygen-dependence. The work highlighted that intracellular or *in vivo* microenvironments, especially those containing thiols, pyridine nucleotides, and unsaturated lipids, are of interest. These likely support chain reactions that amplify oxygen consumption, whereas ascorbate uniquely suppresses these processes. They also confirmed that many of these exhibit dose-rate dependence. Overall, the findings underscore that radiolytic oxygen consumption is highly context-dependent, chemically complex, and not universally high enough to explain FLASH effects, but may be significant in specific intracellular environments where reducing agents and radical-propagation pathways dominate. It is conceivable that certain normal tissues have radical-scavenging capabilities that resist oxygen-related damage at high dose rates, whereas tumor tissue does not exhibit as dominant a response.

Subsequently, several *in vivo* studies have measured the magnitude of oxygen consumption during UHDR irradiation under varying oxygen levels. The *in vivo* results of these studies are summarized in the image below:

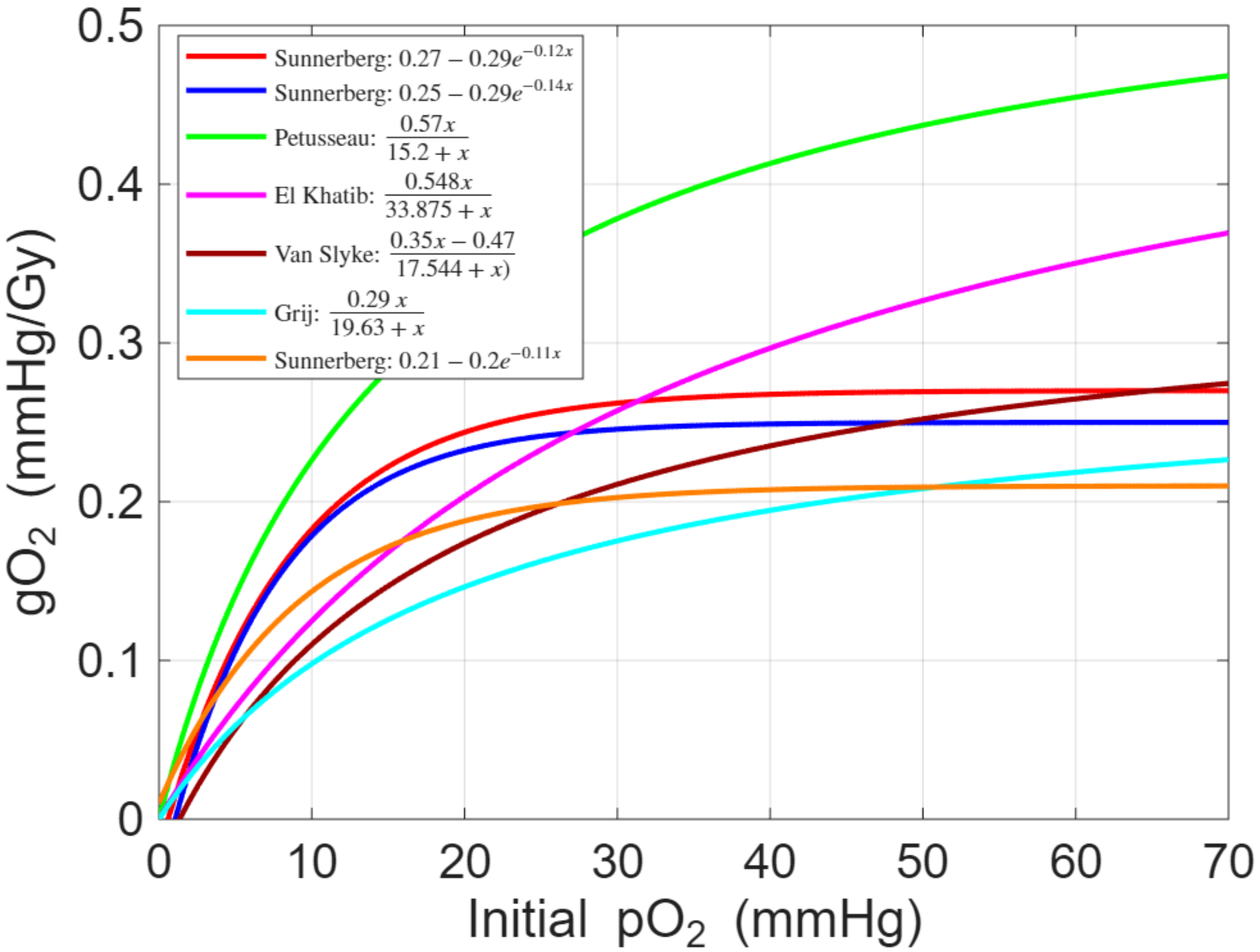

**Figure 6.** A summary of *in vivo* oxygen g-value measurements during UHDR irradiation is plotted to show the established dependence upon initial pO2 value. The lines show results from multiple studies [5,72,74–77,81].

These studies cover electron and proton radiation, intracellular and extracellular oxygen measurements. All studies were performed in mice, with most measuring in murine skin. There is high similarity in both the shape and magnitude among all of these *in vivo* curves. They also closely resemble in-vitro results. It appears that oxygen consumption is directly dependent on initial oxygen concentration at low values; it then reaches a plateau after an intermediate oxygen concentration, near $pO_2 \approx$ 15-25 mmHg. It is also worth noting that current experimental measurements of $gO_2$ suggest that this is independent of tissue type [10]. It is not clear whether the lack of tissue dependence was due to the measurement process or is an inherent feature of FLASH oxygen consumption, but further work might elucidate this.

### *3.4 Oxygen Consumption Data Does Not Support Widespread Oxygen Depletion as the Cause of Radiobiological Hypoxia Sparing in FLASH*

One of the major early theories of the FLASH mechanism was that radiolytic oxygen consumption was so rapid in UHDR that it depleted local oxygen, reducing the availability of oxygen for ROS production [17,82]. This theory was supported by older data from the 1970's [83], but would only apply to reduced overall OER and does not explain the selective sparing of normal tissue versus tumor tissue. Indeed, if such an effect were occurring, it might be expected that tumor tissue with a natively low pO2, perhaps near 5-15 mmHg, would be preferentially spared compared with normal tissue, which starts at 20-40 mmHg. However, the g-values observed here are small compared to those in normal tissue, with 0.1-0.4 mmHg being reasonable values. For a 20 Gy dose, this corresponds to a reduction of approximately 2-4 mmHg at maximum, which is well below full tissue oxygen levels. So the concept that UHDR irradiation depletes tissue oxygen to a level of radiobiological hypoxia seems inconsistent with the data, as illustrated in Figure 7. An extensive study of this was conducted by Grilj et al., who showed g-values of 0.13-0.22 mmHg/Gy in tumors, skin, muscle, and brain tissue of mice, and found that these values were largely independent of tissue type. These numbers match expected values and further support the lack of evidence for radiolytic depletion of oxygen to levels associated with radiobiological hypoxia. Their observed plateau for g-values was above 30 mmHg, after which baseline pO2 became independent. Importantly, they demonstrated a temporal trade-off between ROC and oxygen supply, as shown in Figure 4. Importantly, their normal-tissue protection in the brain did not correlate with oxygen consumption or the degree of oxygen depletion, supporting the idea that radiolytic oxygen depletion is not a mechanism underlying FLASH sparing. Nearly all experimental data summarize that, although UHDR consumes oxygen faster than the biological rate of oxygen resupply, it does not deplete oxygen to the point of transient hypoxia. So it seems unlikely that this change would elicit the tissue-sparing effect proposed by the oxygen-depletion hypothesis.

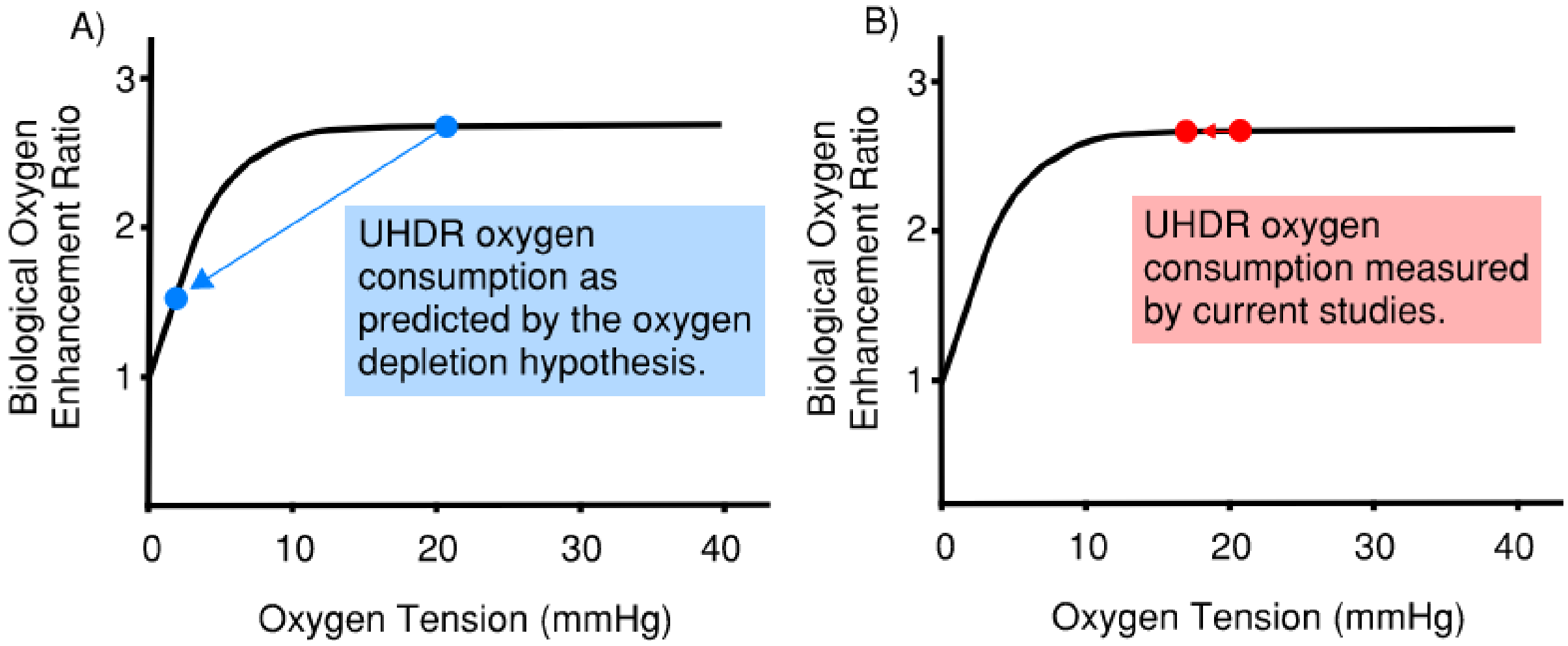

**Figure 7.** Comparison of (A) oxygen consumption predicted by the FLASH oxygen depletion hypothesis and (B) measured FLASH oxygen consumption. Measured oxygen consumption is significantly lower at g-values around 0.25 mmHg/Gy than at the g-value predicted by the oxygen depletion hypothesis (~1 mmHg/Gy).

### *3.5 Saturation of Oxygen Consumption at High Initial Oxygen Levels*

The observed saturation of $gO_2$ above a certain $pO_2$, and the fact that this is independent of tissue type, suggest that it has a Michaelis-Menten-type functional form, where the reaction is initially dependent on sufficient oxygen being available but then becomes saturated and limited by the ROS produced. Within this framework, oxygen acts as an effective $gO_2$ substrate: $gO_2$ increases with increasing oxygen availability at low initial $pO_2$, consistent with a substrate-limited regime. However, beyond a characteristic threshold $pO_2$, the system transitions to a saturation regime in which further increases in oxygen availability no longer significantly increase $gO_2$. Notably, reported threshold pO2 values vary substantially across studies. This variability likely arises from differences in experimental conditions, including radiation modality (e.g., electrons versus protons), beam structure (pulse width, dose rate, and temporal microstructure), and the accuracy and temporal resolution of oxygen measurement techniques.

Mechanistically, primary radicals generated through water radiolysis (e.g., $HO^{\bullet}$, $e_{aq}^{-}$, H•) rapidly interact with molecular oxygen, as well as with major biomolecular targets (lipids, protein, DNA) to form secondary reactive species. A recent study developed and validated a kinetic model of ROC to explain its dependence on radiation dose rate and initial oxygen pressure [84]. This framework indicates that key radiolysis species (e.g., HO•, $e_{aq}^{-}$, $O_2^{•-}$, and $HO_2•$) govern oxygen consumption, with their yields influenced by baseline $pO_2$, dose rate, linear energy transfer (LET), and the chemical environment. Within this model, $gO_2$ follows a functional form analogous to Michaelis–Menten kinetics. At low oxygen levels, the reactions are oxygen-limited, and increasing $pO_2$ enhances the probability of oxygen fixation. At higher oxygen concentrations, $gO_2$ exhibits saturation behavior, analogous to Michaelis–Menten kinetics, in which the reaction rate is limited by system turnover rather than by substrate availability. Collectively, this framework links radiation chemistry to oxygen dynamics in FLASH radiotherapy and helps explain how dose rate and initial oxygen levels modulate ROC.

This saturation behavior has important implications for interpreting FLASH effects, as it indicates that under well-oxygenated conditions, ROC cannot increase indefinitely as $pO_2$ increases. Consequently, oxygen depletion alone may not fully account for differential tissue sparing, particularly in regions with high baseline oxygenation, where ROS-driven kinetics impose an upper bound on oxygen consumption.

| **Key Points** | |
|---|---|
| **1** | Oxygen consumption ($\Delta pO_2$) at low $pO_2$ is directly linked to baseline oxygen, following first-order reaction kinetics (i.e., $O2 + R\cdot \rightarrow ROO\cdot$ ), but above about 20 mmHg, the consumption rate saturates. g-values are $\Delta pO2$ per unit dose. |
| **2** | $\Delta pO_2$ appears to be dominated by peroxyl formation (ROO·) but is not exclusive to this. |
| **3** | G-value decreases with increasing dose rate in both protons and electrons. |
| **4** | G-values range between 0.1 and 0.5 mmHg/Gy, depending on the measurement medium or tissue. |

| **Open Questions and Future Directions** | |
|---|---|
| **1** | Could micro-local heterogeneity of oxygen be present and be related to $DpO_2$ change and radiobiological damage changes? |
| **2** | Could radical scavengers dominate the DpO2 changes between tissue types and subsequently drive the observed FLASH effect? |
| **3** | Could the gO2 vs pO2 curve be used to estimate the concentration of ROS produced in a tissue type, and the balance point between produced versus scavenged radicals? |
| **4** | Is intracellular oxygen near DNA different than extracellular oxygen, and does the UHDR ROS cascade differ there? |

## 4.0 The Link Between Oxygen and FLASH Biological Tissue Sparing

**Table 3.** FLASH studies of the effects of inhaled oxygen modulation on FLASH outcomes.

| Study | Radiation Type | Model | Assay | $pO_2$ Mod Technique | $pO_2$ Measured (Y/N) and Conclusions | sparing (Y/N) | ROC (Y/N) | Conclusion |
|---|---|---|---|---|---|---|---|---|
| Leavitt et al. (2024) | Electrons (6 MeV) | *In vivo*: Murine Lung, GBM, H&N Tumors | Survival: Tumor growth kinetics, RNAseq, Pimonidazole | Clamping, Room Air, Carbogen | No $pO_2$ monitoring OxyLED. Small differences between clamped and carbogen, limited group sample size (n=3). Large difference between clamped and normal/carbogen | No | No | UHDR retains tumor-killing efficacy in hypoxic tumors, whereas CDR does not. Support for a differential FLASH response in tumors vs normal tissue |
| Karsch et al. (2022) | Electrons (30 MeV)<br>Protons (224 MeV) | *In vivo*: Zebra fish embryos | Survival: Morphology and Development | Dissolved oxygen modulation via irradiation time post-sealing | Yes, $pO_2$ with OxyLite. Rediffusion of oxygen is not present in UHDR. Single reference value | Yes | No | Volume restriction, oxygen consumption, and environmental conditions did not alter embryo well-being and radiation response. Irradiating < 10 mmHg demonstrated a FLASH with protons. |
| Pawelke et al. (2021) | Electrons (30 MeV) | *In vivo*: Zebra fish embryos | Survival: Morphology and Development | Dissolved oxygen modulation via irradiation time | No $pO_2$ monitoring No $\Delta pO_2$ monitoring OxyLite | Yes | No | Observed reduction in overall damage to UHDR cohorts in low pO2 environments as compared to CDR |
| Adrian et al. (2021) | Electrons (10 MeV) | In-Vitro human cancers: breast (MCF7, MDA-MB-231) cervical (HeLa EP, HP-subclone), colon (WiDr), SCC (LU-HNSCC4) | Survival: Clonogenic, DNA-double strand break foci, Cell cycle analysis | None, Normoxic pO2 (21% O2) | No $pO_2$ monitoring No $\Delta pO_2$ monitoring Ambient 21% oxygen maintained during irradiation | Yes | No | In vitro, FLASH sparing was observed in some cell lines, indicating a possible biological susceptibility to FLASH. |
| Adrian et al. (2020) | Electrons (10 MeV) | In-Vitro: Prostate cancer cells (DU145) | Survival: Clonogenic | Irradiated from 1.6-20% | Yes, $pO_2$ electrode. No $\Delta pO_2$ monitoring Hypoxia chamber used to lower $pO_2$ at the time of irradiation. | Yes | No | Hypoxic (1.6% O2) conditions produced a FLASH effect, which gradually diminished as pO2 increased to 20%. |
| Nias et al. (1970) | Electrons (10 MeV) | In-Vitro: Cervical cancer (HeLa) | Survival: Clonogenic | Sealed in vials (cellular respiration induced hypoxia) | No $pO_2$ monitoring No $\Delta pO_2$ monitoring Hypoxia assumed via cellular respiration dynamics. | Yes | No | FLASH sparing observed compared to 1Gy/min in hypoxic cells vs. aerated cells. |
| Cao et al. (2021) | Electrons (10 MeV) | In-Vitro: BSA<br>*In vivo*: Nude mice, Human breast MDA-MB-231 cells | None | None | Yes $pO_2$ with Oxyphor<br>Yes $\Delta pO_2$ measured. Oxyphor and electrodes.<br>Showed first transient decrease in pO2 during FLASH, no change during CONV. | N/A | Yes (negative conclusion) | ROC is likely not large enough to induce transient hypoxia.<br>Lower $\Delta pO_2$ observed in UHDR vs CDR *in vitro*.<br>Lower oxygen consumption observed in tumors than normal tissue. |
| Montay-Gruel et al. (2019) | Electrons (6 MeV) | *In vivo*: Female B6 Mice | Novel object recognition, Discrimination index | Carbogen with Iso | No $pO_2$ monitoring<br>No $\Delta pO_2$ monitoring<br>Carbogen decreases FLASH sparing, equivalent CDR damage maintained. | Yes | No | Increasing oxygen tension within the brain through carbogen breathing reversed the FLASH effect. No changes in CDR observed under carbogen vs. RA. |
| Hansen et al. (2025) | Electrons (16 MeV) | *In vivo*: Normal female C3H Mice | Skin damage assay | Vascular clamping | No $pO_2$ monitoring<br>No $\Delta pO_2$ monitoring<br>Tissue pO2 measured in a separate cohort, not matching irradiation setup. | Yes | Yes | Clamping reduced overall damage; hypoxia induced equivalent damage in FLASH and CDR. Supports oxygen consumption as mechanism. |
| Dewey et al. (1959) | Photons (5MV) | In-Vitro: Bacteria, Serratia Marcescens | Survival: Clonogenic | Dissolved gas modification | No $pO_2$ monitoring<br>No $\Delta pO_2$ monitoring | Yes | Yes (not directly measured) | Irradiating bacteria under hypoxic conditions elicits a FLASH sparing effect not seen under normoxic conditions. |
| Zhang et al. (2023) | Protons (230 MeV) | *In vivo*: Normal female FVB/N Mice | Skin contracture assay | Ketamine xylazine, normoxic (room air), hypoxia (clamping), 100% O2 cohorts. | No $pO_2$ monitoring<br>No $\Delta pO_2$ monitoring<br><br>No FLASH sparing in clamped cohort.<br><br>No FLASH sparing in 100% O2 cohort. | Yes | Potential | Unresolved ROD hypothesis. May be true for select tissues. Potential support for ROD hypothesis due to disappearance of FLASH sparing at high pO2. |
| Iturri et al. (2023) | Protons (227 MeV) | *In vivo*: Fischer 344 Rats Naive and with RG2 Glioma | Neuro-behavioral, MRI assessment of brain injury. | 2.5% Iso + 21% O2 or 70% O2 | No $pO_2$ monitoring<br>No $\Delta pO_2$ monitoring<br>Oxygen measured in skin, not brain.<br>No oxidative parameters measured. | Yes | No | Damage differential between UHDR and CDR maximized with room air anesthesia—higher damage observed under 70% O2 supplementation. |
| Kasamatsu et al. (2025) | Carbon Ions (400 MeV/u) | *In vivo*: Normal female C3H Mice | Survival: NTCP, crypt counting, Weight loss | None | No $pO_2$ monitoring<br>No $\Delta pO_2$ monitoring | No | No | High LET irradiation of the GI at the dose level studied did not produce a statistically significant FLASH sparing effect. |
| Sesink et al. (2025) | Electrons (9 MeV, 6 MeV) | *In vivo*: Normal female B6 Mice (TAI) and female BALB/C (Skin) | Survival: Weight loss assay (GI tox criteria), skin tox assay | Ketamine-xylazine supplemented with 21% O2 or >90% O2 | No $pO_2$ monitoring<br>No $\Delta pO_2$ monitoring<br>Pre-irradiation oxygen tension verified in a separate cohort using OxyLED. | Yes | No | Oxygen supplementation eliminated FLASH sparing in the intestine, but not in skin. |

### *4.1 Inhaled Oxygen Modulation: Air, Oxygen, Carbogen*

Modifying the initial pO2 at the irradiation site modulates the in vivo differential between UHDR and CDR irradiations. Multiple studies have attempted to alter the observed FLASH effect by increasing the oxygen content of inhaled anesthetic gas, using either carbogen or 100% oxygen, with the goal of enhancing tissue $pO_2$.

In an early study by Montay-Gruel et al., the neurocognitive protective differential between UHDR and CDR electron irradiations was diminished under carbogen (95% O2, 5% $CO_2$) anesthesia breathing conditions [85]. It has been well established that carbogen increases oxygen within cerebral tissues, raising pO2 above 75 mmHg [86], and that this increase in pO2, as Montay-Gruel hypothesized, actually decreased the sparing effect of UHDR versus CDR at isodose levels. Similarly, Iturri et al. demonstrated in mice skin and brain irradiated using protons under 100% oxygen with isoflurane anesthetic, that the UHDR sparing effect was suppressed when compared to outcomes under 21% $O_2$ (room air equivalent) gas supplementation given a total dose of 25 and 15 Gy delivered at ~250 Gy/s [3]. Differences were observed in the severity of dermatitis in the irradiated region (murine head), duration of radiodermatitis, and neural radionecrosis of the hippocampus between the UHDR with 100% O2 and the UHDR with 21% $O_2$ cohorts. Behavioral changes were also observed between these two cohorts, with permanent mnemonic alterations found in the 100% UHDR O2 cohort, but not in the 21% UHDR O2 cohort, indicating more serious damage with full oxygen.

Physiological and experimental factors, particularly anesthesia, further modulate tissue oxygenation and influence FLASH outcomes. Tavakkoli et al. (2023 demonstrated that sex differences in tissue oxygenation exist in mice when anesthetized using isoflurane, with significant differences in skin damage apparent under a 100% $O_2$ gas mixture [94]. Additionally, differences in achieved $pO_2$ values have been observed in the GI by Tavakkoli et al. 2025 when the choice of anesthesia is changed. Under ketamine/xylazine injections at a dose of 20 mg/kg, profound hypoxia is induced in the gastrointestinal (GI) model when compared to isoflurane at 1.5% and a room air gas mixture [95]. The modification of anesthesia also altered the presentation of the UHDR sparing effect in the GI, as manifested in both weight-loss and survival assays of mice irradiated at a determined sub-lethal dose, with only ketamine/xylazine UHDR mice showing significant sparing.

Despite these observations, an important consideration is the absence of real-time tissue pO2 measurements during irradiation, which is needed to fully quantify the implications of inspired oxygen on the tissue being irradiated and any depletion of oxygen during UHDR irradiations [2,91–93]. In addition to the complexity posed by anesthesia conditions, there is the interplay between LET and oxygen consumption. Recent publications on carbon ion irradiation have produced two different findings regarding the damage induced by this. In Kasamatsu et al. (2025, UHDR carbon ion irradiations were found not to preferentially spare GI function over CDR carbon ion irradiation [96]. No significant differences were observed at a sublethal dose level in weight loss, survival, number of apoptotic crypts, apoptotic cells per crypt, and crypts remaining post-irradiation between the UHDR and CDR cohorts. Notably, the anesthetic used in this experiment was secobarbitol, which has not been studied for its effects on tissue oxygenation during UHDR deliveries.

While oxygen modulation has been readily shown to alter the presentation of a UHDR normal tissue-sparing effect, this is only half of the definition of FLASH, and efforts demonstrating the same tumor control while modulating tumor oxygenation have been relatively sparse. Multiple FLASH studies have shown FLASH effect–equivalent tumor control between UHDR and CDR [87–90] under single-oxygenation conditions. However, comprehensive studies of oxygen modulation on tumor control in orthotopic, heterotopic, or spontaneous tumors are lacking. This issue of tumor control in FLASH is a critically important area of FLASH radiation biology that has been comparatively understudied versus the normal tissue response. This issue is likely due in part to the complexity of quantifying response amid high inter-animal variability, and in part to the fact that few studies have shown differences in tumor response across varying dose rates. However, this is an area that warrants significantly more examination, perhaps with tumors in which oxygenation can be systematically varied or with phenotypic models to better understand why FLASH does not affect tumor response. Overall, current evidence indicates that increased tissue oxygenation tends to reduce or eliminate the FLASH normal tissue sparing effect. However, the lack of quantitative $pO_2$ measurements and the influence of confounding experimental variables underscore the need for more controlled and mechanistically driven studies. Table 3 summarizes recent *in vivo* studies investigating the effects of inhaled oxygen modulation on FLASH outcomes across different tissues, measurement techniques, and radiation modalities.

### *4.2 The FLASH Effect in Hypoxic Tissue*

Early analysis of the interplay between diffusion and hypoxia in the UHDR sparing effect showed that, in a naturally hypoxic animal model (the murine tail), the total dose required to produce necrosis increased when total delivery times were kept below 4.5 seconds, a time deemed to match early estimates of oxygen diffusion timelines [97]. Under these conditions, early *in vivo* analysis of UHDR tissue sparing was observed, with a link between re-oxygenation and a UHDR sparing effect in electron FLASH. Sunnerberg et al. confirmed this finding using a separate method to modulate the total duration of a UHDR delivery. When the interval between two 12.5 Gy deliveries (totaling 25 Gy in a single treatment) was varied, UHDR sparing was preserved as long as the total treatment time remained under 5 seconds. Using real-time oximetry, the diffusion rate was mapped during irradiations, and it was found that when the total delivery period exceeded the time required for full re-oxygenation, the UHDR sparing effect disappeared [91].

These findings have been replicated using vascular clamping, as shown by Hansen et al. (2025). Using an optical oxygen reporter, the presence of hypoxic and normoxic conditions was confirmed in clamped and unclamped murine legs before irradiation. Key findings from the study showed that the normal tissue-sparing effect disappeared across multiple dose levels in the clamped cohort, whereas the unclamped cohort demonstrated a significant sparing effect. It was also shown that the dose required to produce a fixed level of skin damage increased under the clamped condition, closely aligned with a conventional understanding of the oxygen enhancement ratio [2]. While the paper is notable for showing that the UHDR sparing effect disappears at low tissue pO2 values and further supports the hypothesis that the UHDR sparing effect exists within a fixed window of $pO_2$ values, the oxygen values reported in the paper are not those at the time of irradiation, nor is consumption during irradiation measured. It is also worth noting that the oxygen values reported in the paper may not be representative of those at the time of irradiation, as the oxygen measurements appear to have been taken under room-temperature air conditions. In contrast, irradiation was performed with the mouse leg submerged in 25℃ water, a temperature that may itself lower tissue pO2 (as evidenced by the relatively large doses required to induce ulceration).

These early *in vivo* studies of oxygen consumption and damage from dose-rate variation led to the hypothesis that there is a biological “window” of pO2 values within which irradiation effectively increases the differential response between CDR- and UHDR-exposed tissues. To further assess the impact, measurement, and alteration of pO2 at the time of irradiation have been explored through both vascular compression, which induces anoxia, and modification of the inspired oxygen level (FiO2) or the type of breathing gas (medical air, 100% O2, carbogen, or anywhere in between). Hunter et al. (2026) described the effects of extensive modulation of inhaled anesthetic breathing gases and limb compression on the presentation of a UHDR skin-sparing effect at a fixed dose level. It was observed that at a known ulceration-producing dose level, the UHDR sparing effect is preserved only under mild hypoxia, disappearing under complete hypoxia and normoxic-to-hyperoxemic conditions. It was additionally observed that progression to ulceration persisted predominantly in UHDR mice with tissue oxygen tensions above 16 mmHg, and was less observed in tissues with tensions below 15 mmHg [6].

Perhaps the most notable tumor response study using clamping for anoxia was conducted by Leavitt et al., who demonstrated that while CDR tumor damage was limited by severe hypoxia, clamping the tumor did not alter UHDR damage in tumors regardless of clamping-induced hypoxia. This observation was interpreted to suggest that UHDR tumor damage is independent of hypoxia, representing a radical change in the mechanism of radiation response relative to CDR, with essentially no OER effect. Molecular response studies in that work indicated that UHDR irradiation drove stronger cell-cycle shutdown and ribosomal arrest under clamped hypoxia, indicating less potential for repair. They proposed that UHDR irradiation induces a hypoxia-inducible factor-like metabolic response that creates this hypoxia-resistant damage mechanism. This study focused on tumor response, so response variations between normal tissue sparing and tumor tissue were not directly compared; however, the finding that UHDR induces tumor death independent of hypoxia is highly notable. It is important to note, however, that the authors demonstrated a differential response between UHDR cohorts under carbogen supplementation, with no significant differences in tumor size doubling times between carbogen-UHDR and carbogen-CDR groups [98].

### *4.3 In vitro Oxygen Modulation Studies*

The overall status of interpreting the FLASH effect *in vitro* has, put simply, been confusing. Early studies by Cygler et al. (1994) showed no differences in cell survival curves between UHDR and CDR treatments in glioma and melanoma cell lines [99]. They modulated oxygenation between complete anoxia (98% N2 and 2% CO2) and aerobic (room air) conditions. When these were studied at multiple dose levels, no difference was observed between the UHDR and CDR cohorts. However, it is important to note that the UHDR irradiations used a 20 MeV LINAC, whereas the CDR cohorts were irradiated using a cobalt-60 irradiator. In contrast, more recent findings [100] have indicated that a UHDR-sparing effect exists under normoxic conditions *in vitro*. Adrian et al. (2020) demonstrated that under “normoxic” conditions, no differences between FLASH and CDR-irradiated prostate cancer cells were found. However, when cells were irradiated at varying inspired oxygen concentrations (1.6%- 4.4% O2), the UHDR sparing effect, as assessed by a clonogenic assay, was present and decreased as the oxygen concentration was increased to 8.3%.

Adrian et al. (2021) reported a sparing effect in several cell lines irradiated under normoxic conditions [101]. Statistically significant sparing (associated with dose-modifying factors) was observed in clonogenic assays under 20% O2 for the HeLa (human cervical cancer; both early-passage and subclone models), MCF7 (human breast cancer), and LU-HNSCC4 (squamous cell carcinoma) cell lines, with dose-modifying factors of 1.1-1.32. Additional sparing has been observed in spheroid models under hypoxic conditions in the spheroid core [102]. Although conflicting in some cell models, with some reporting no sparing under normoxia and others finding sparing, it is clear that some effect should exist to explain the variation *in vitro* [99–102]. One possible explanation for the difference between UHDR and CDR *in vitro* is the preferential sparing of mitochondrial function, as outlined by Guo et al. (2022). Under normoxic conditions (21% $O_2$), statistically significant differences in cell survival were observed in skin fibroblasts, with a significant increase in Drp-1 protein expression (a protein that mediates mitochondrial fission) observed only in the CDR cohort [103]. With the introduction of Mdivi-1, a Drp-1 inhibitor, Guo et al. (2022) also showed that mitochondrial fission was reduced under CDR, strengthening the UHDR mitochondrial-sparing argument.

These results paint a complex picture of the state of FLASH *in vitro*, with apparent differences readily observed in some cell lines, though not in others. Differences in oxygen supplementation setups, cell passage numbers, real-time oxygenation measurement techniques, and irradiation setups may account for some of the differences. However, further evaluation of in vitro models is necessary to reproduce the consistency of findings observed in *in vivo* models. Studies that incorporate direct intracellular measurement of oxygen consumption during UHDR irradiations may be beneficial [104].

### *4.4 Interpreting the Link Between Oxygen and FLASH*

Interpreting the range of studies on oxygen and FLASH normal tissue sparing has been challenging; however, there appears to be consensus that baseline normal tissue pO2 values influence the incidence of radiation-induced toxicity and that the oxygen enhancement ratio differs between UHDR and CDR. Data on direct oxygen modulation of the FLASH effect are shown in Figure 2b. The interpretation of this effect and the underlying mechanism remain unclear. Still, most data indicate that in radiation-induced skin toxicity, the toxicity-sparing effect is most pronounced at low pO2 values and diminishes as $pO_2$ rises. However, the FLASH sparing effect appears to disappear under anoxic conditions, where the apparent OER-type curve shifts to the right at UHDR and eventually converges to the conventional equivalent above 20 mmHg. One account of this effect has been demonstrated by Hunter et al. 2026 [116], which illustrates an equivalent effect obtained via an *in vivo* assay of skin toxicity. The differential between tumors under different oxygenation conditions (21% $O_2$, vascular compression, and Carbogen) has also shown equivalent tumoricidal efficacy of FLASH between hypoxic and normoxic tumors, indicating that the OER may play a separate role in the treatment of conventional tumors [98]. While these two studies indicate some interplay between oxygen and the presentation of the FLASH effect, including both independent differentials between tumor oxygenation conditions and normal tissue oxygenation conditions, it is clear that evaluating both normal tissue and tumors in a single experiment is necessary to establish a paired evaluation of oxygen's role under UHDR conditions.

| **Key Points** | |
|---|---|
| **1** | Hyperoxia and anoxia reduce the presentation of UHDR toxicity sparing to normal tissues. |
| **2** | Tumor treatment efficacy **<u>may</u>** be unaffected by hypoxia or anoxia under UHDR |
| **3** | In-Vitro studies can be conducted to show a UHDR sparing effect when oxygen is modulated to a hypoxic level |
| **4** | Multiple studies claim to correlate oxygen with a UHDR sparing effect, though most do not have concurrent irradiation and oxygen measurements. |

| **Open Questions and Future Directions** | |
|---|---|
| **1** | Is the alteration of tumor damage and normal tissue toxicity by UHDR caused by a cascade of ROS *in vivo*, or is an independent mechanism at play? |
| **2** | Simultaneous measurement of tissue $pO_2$ at the time of irradiation is necessary to correlate normal tissue toxicity or tumor effects with oxygen modulation conditions. |
| **3** | Can we separate DNA damage caused by direct effects from that caused by indirect effects to understand the mechanism of FLASH better? |

**5.0 Photophysics of Oxygen Consumption and Radiation Damage**

The initial radiolytic event occurs on a femtosecond timescale, as the incident radiation excites and ionizes $H_2O$. The ionized water radical cation ($H_2O^{+\bullet}$) is extremely short-lived and undergoes rapid proton transfer to a neighboring water molecule, generating the hydroxyl radical (•OH). The ejected electron rapidly thermalizes by generating a hydrated electron. The cascade of oxygen interactions that follows spans many orders of magnitude in time. Primary ionization and excitation events occur within $10^{-15}$ s, while the hydrated electron is fully formed by ~$10^{-12}$ s. Hydroxyl radical reactions with biomolecules occur on the nanosecond-to-microsecond timescale, while secondary processes such as the Fenton reaction, lipid peroxidation chain reactions, and activation of enzymatic ROS-generating systems (e.g., NADPH oxidase) may persist for minutes to hours post-irradiation.

The observation of ROS is notoriously difficult due to the extremely short lifetimes of key species, such as hydroxyl radicals, which are on the order of nanoseconds in *in vitro* or *in vivo* environments dominated by large ambient protein molecules. The field of ROS measurement is of key importance in biological research due to the central role of ROS in signaling via redox pathways, inflammation, and treatment response [105]. Developed assays to directly measure the ROS cascade include (AR, luminol, PCL-1) [106], identifying a species with a substantial optical signature that is observable [107,108], or using an EPR signature [53,109]. There are myriad pitfalls and issues with commonly used *in vitro* and *in vivo* quantitative assays [110]. Luminescent reporters lack specificity and face diffusion limitations when examining the most important, rapidly interacting molecules, such as hydroxyl radicals. The only molecule in the ROS pathway with an optical signature is the absorbance due to aqueous electrons, which has yet to be studied in a non-pure-water environment [107]. EPR probes can observe fast-decaying molecules, but in complex systems with high noise levels, it is difficult to couple them to radiation sources. With these in mind, a key molecule involved in the ROS pathway is oxygen. Given its crucial role in biology, measuring tissue oxygenation in vivo offers a wealth of opportunities for continued exploration.

Oxygen is a key molecule in the chemical process of radiolysis. It is one of the most significant modulating parameters for radiation-induced damage via the OER, with its own unique damage type: oxidative fixation. On a chemical level, oxygen is consumed primarily through interactions with hydrogen radicals (eq 1), $eaq^-$ (eq 2), and proteins (eq 3), which result in final products that generate damage from hydrogen peroxide, superoxide, or peroxyl radicals. This chemistry is key to understanding the radiosensitizing effect of oxygen, as interactions with protein radicals generate peroxyl radicals, which are thought to be the primary contributors to the OER effect [29]. For example, Cao et al. quantified the transient production yield of $e_{aq}^-$ in water under UHDR [111]. The study demonstrated an increase in the g-value of $e_{aq}^-$ at higher instantaneous dose rates. As illustrated in Figure 9, the increase in $e_{aq}^-$ has important downstream effects on $H_2O_2$ formation and may contribute to a reduction in overall DNA damage during UHDR irradiation. While the reaction of oxygen and protein radicals is not nearly as diffusion limited as its reaction with hydrogen radicals or $e_{aq}^-$, the overall concentration of protein substrates is substantial *in vitro* *and in vivo* causing an up to 5x gain in oxygen consumption when comparing pure water to increasing protein solution demonstrated by Sunnerberg et al. [77], and, depending on the complexity of the environment, oxygen consumption can vary substantially [80].

$$H\bullet + O_2 \rightarrow HO_2\bullet \qquad k = \sim 10^{10} \tag{1}$$

$$e_{aq}^{-} + O_2 \rightarrow O_2\bullet^{-} \qquad k = \sim 10^{10} \tag{2}$$

$$R\bullet + O_2 \rightarrow ROO\bullet \qquad k = 10^{7} - 10^{8} \tag{3}$$

Due to the difficulty and complexity of measuring these molecules, various Monte Carlo and numerical rate equation calculations were used to identify perturbations in non-homogeneous stage chemistry [112–115]. When analyzing changes in oxygen consumption, “simple” numerical models, such as those of Tan et al. [113], are in good agreement with both *in vitro* and *in vivo* experimental data [1,2]. Tan et al. have noted that peroxyl and superoxide radicals undergo significant dose-rate-induced changes, both of which are products of oxygen consumption, as noted in equations 2 and 3. The dose-rate dependence of the overall time-integrated peroxyl radical is approximately linear above 30 Gy/s, with only a minor decrease in overall peroxyl production. In comparison, superoxide decreases sharply by a factor of 3 between conventional dose rates and 30 Gy/s, then plateaus.

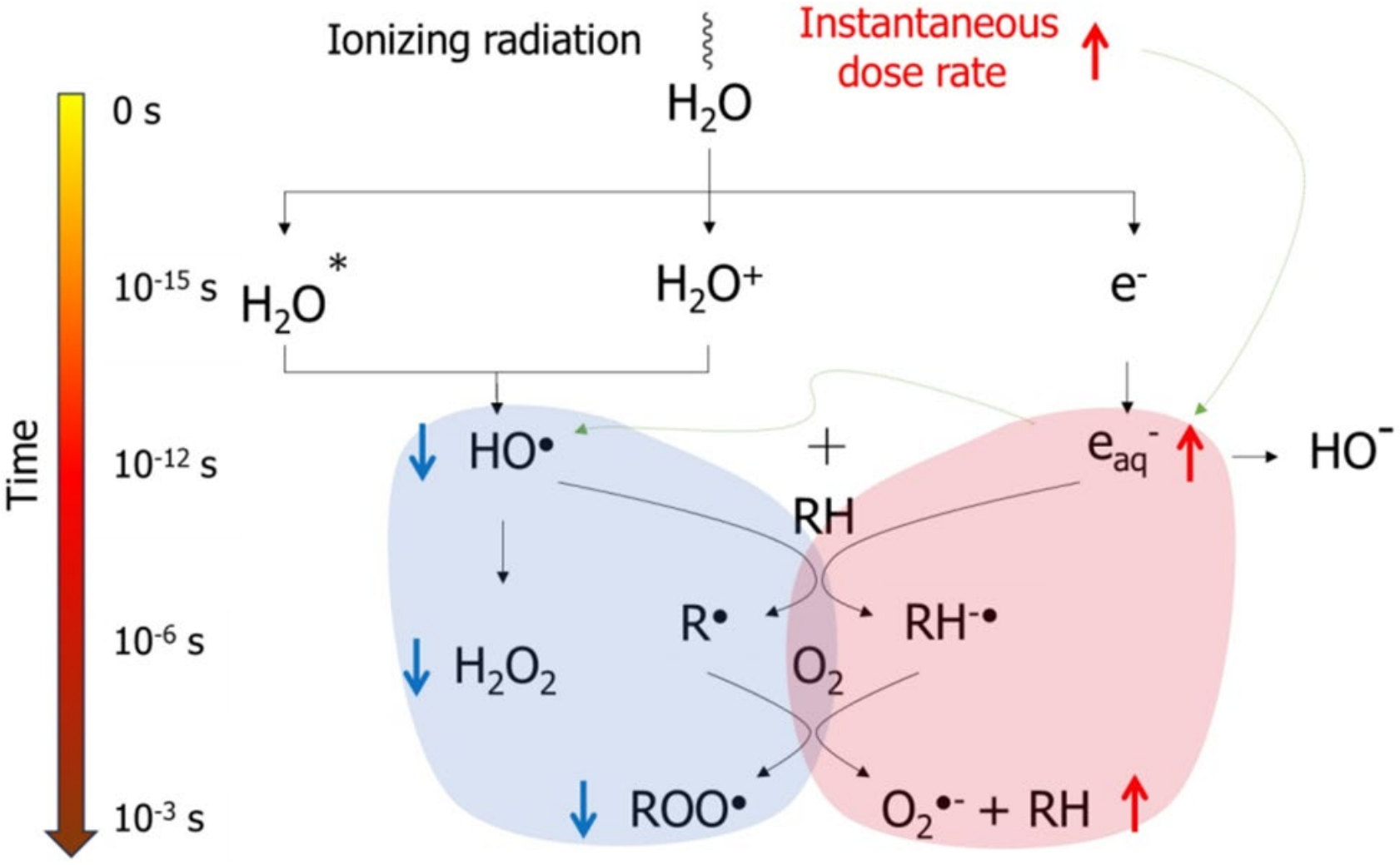

**Figure 9.** Schematic illustration of some of the key influential products from hydrolysis as a function of timescale and chosen based upon their relevance to oxygen and possible subsequent biological toxicity [111].

| Key Points | |
|---|---|
| **1** | Accurate measurement of nearly all highly damaging ROS produced *in vivo* is not possible, making $\Delta pO_2$ measurement one of the few means of assessing transient radiation chemistry in vivo. |
| **2** | Oxygen consumption and production occur at many stages in the transient cascade of ROS. |
| **3** | Local microenvironment alters the ROC through multiple mechanisms, such as increased peroxyl formation in consumption, ROS scavenging, and oxygen regeneration. |
| **4** | High concentration of proteins and the mobility of oxygen and radicals/ions make ions and proteins the most likely interactions, although the mobility of large proteins, lipids, and DNA is very low compared to O2 and ions. |

| **Open Questions and Future Directions** | |
|---|---|
| **1** | Can the nature of inhomogeneous and homogeneous radiation-chemistry cascades really be simulated to predict *in vivo* changes? |
| **2** | Why does the oxygen dependence vary with tissue type? |
| **3** | Is it the hypoxia or microchemical environment of tumors that leads them to have no FLASH sparing? |
| **4** | Do scavengers influence the magnitude of the FLASH effect by modulating both ROC and the downstream tissue response? |
| **5** | Is there direct proof that oxygen depletion or initial $pO_2$ causally links to tissue sparing? Or is all evidence correlative? |

**Conclusions**

The FLASH effect in radiotherapy offers the potential to widen the therapeutic window by sparing normal tissue while maintaining tumor control. Despite rapid progress, the underlying mechanisms remain incompletely resolved; this review focuses on the role of oxygen dynamics, integrating evidence from *in vivo* measurements, oxygen modulation studies, and radiation chemistry. There are many steps in the process, from the initial radiolysis of water to the damage and tissue sparing of FLASH, in which oxygen and oxygen-related factors influence what occurs. An illustration of this is shown in Figure 10, with several of the more dominant known factors affecting FLASH (annotated with arrows to indicate their effect on sparing) and several hypothesized factors involved in the mechanism.

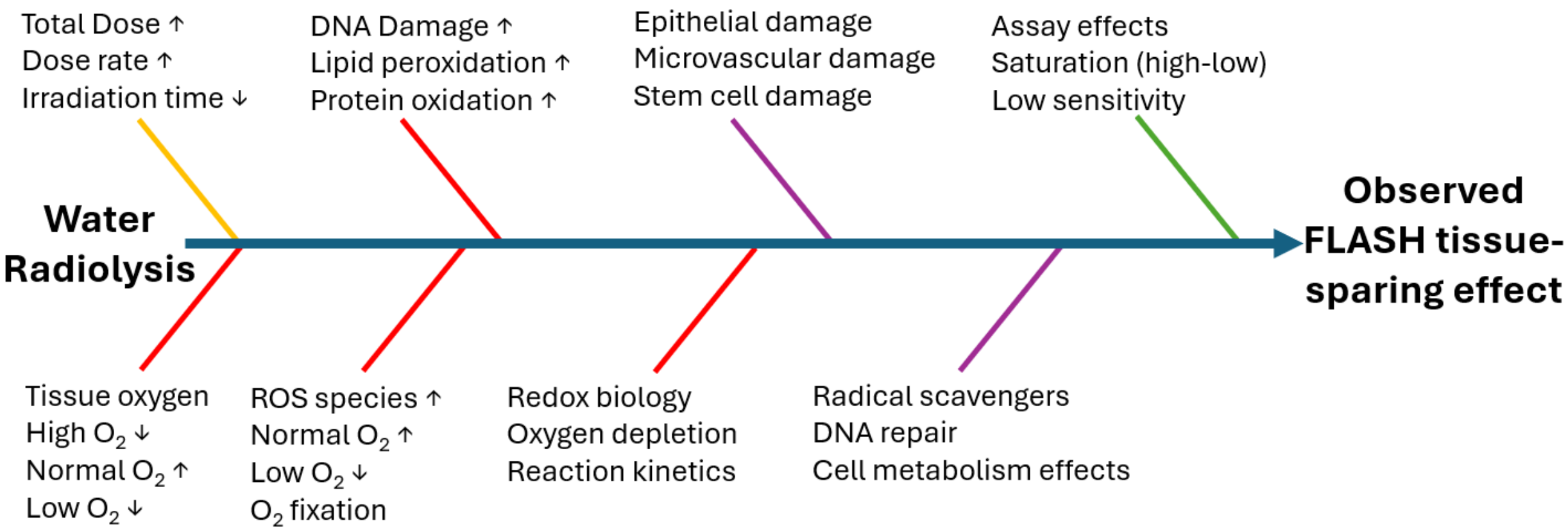

**Figure 10.** A fishbone process diagram shows known or hypothesized factors that influence the FLASH sparing effect, with oxygen-related inputs shown in red and downstream oxygen-related parameters in purple. This flow matches the rough timeline from initial water radiolysis through to the biological observation of reduced damage or sparing.

On the topic of oxygen supply and depletion during irradiation, current data indicate that rapid ROC definitively occurs during UHDR irradiation; however, the magnitude of oxygen depletion appears generally insufficient to induce broad tissue radiobiological hypoxia in most tissues. Instead, the FLASH effect appears to be strongly modulated by baseline tissue oxygenation, with multiple studies on skin toxicity suggesting a limited $pO_2$ window in which normal tissue sparing is maximized. Experimental manipulation of oxygen levels through inhaled gases, vascular interventions, and anesthesia further supports the sensitivity of FLASH responses to physiological and experimental conditions. However, in contrast to all this, the timed delivery of pulsed radiation with a split-dose scheme [5] showed that FLASH sparing was preserved when timing or dose rate was adjusted to allow 5-10 seconds of delivery, a timescale that matches oxygen diffusion well. This discrepancy between quantitative estimates of depletion and the timescale of the FLASH observation remains unresolved. Ultimately, these data are also limited in their generality, primarily because they are largely skin data, and further exploration in other tissues is difficult due to the inability to modulate oxygen in most vital organs effectively.

At the same time, advances in radiation chemistry, both experimental and simulation-based, demonstrate that UHDR irradiation alters radical and ROS production pathways, apparently shifting biological effects away from hydroxyl radical–dominated damage toward greater contributions from solvated electrons and downstream related ROS species. It seems likely that this shift in ROS, along with the associated scavenging and repair mechanisms present in normal tissues, is involved in the shift in toxicity response from CDR to UHDR irradiation. These findings suggest that the FLASH effect is unlikely to be explained by a single mechanism but rather arises from a complex interplay among oxygen availability, dose-rate–dependent chemical processes, and tissue-specific biological responses. Further definition of this with computational studies and/or basic analytical chemistry experiments is critically needed, with codes that match experimental data, or are used to drive hypotheses for testing with appropriate experiments.

Looking forward, several key directions are critical for advancing the field and unraveling the role of oxygen in radiation response and how it is altered in UHDR irradiation. These include:

- Development and implementation of high-resolution, real-time oxygen-sensing technologies will be essential to quantify transient oxygen dynamics during UHDR irradiation directly.
- Systematic studies that independently vary baseline $pO_2$, dose per pulse, and instantaneous dose rate are needed to decouple the relative contributions of oxygen depletion and radiation chemistry.
- Integrating experimental measurements with multiscale computational modeling of radiolysis, oxygen diffusion, and biological response may provide a more comprehensive mechanistic understanding.
- Greater emphasis should be placed on standardized reporting of irradiation parameters and physiological conditions, including anesthesia, temperature, and tissue-specific oxygenation, to improve reproducibility across studies.
- Expanding investigations into tumor models under controlled oxygenation conditions will be essential to determine whether the oxygen dependence of normal tissue sparing extends to tumor response.

Ultimately, bridging the current gap between oxygen measurement, radiation chemistry, and biological outcome will be crucial for optimizing FLASH protocols and enabling safe and effective clinical translation.

## Conflict of interest

Brian Pogue declares financial involvement with DoseOptics LLC, and Harold Swartz declares financial interest in Clin-EPR LLC. The authors declare no other conflicts of interest.

## Acknowledgments

This work has been funded by NIH Grants U01 CA260446 and R01 CA271330, and by the shared services of the Dartmouth Cancer Center P30 CA023108.